\documentclass[aps,prb,twocolumn,amsmath,amssymb,superscriptaddress,reprint,longbibliography]{revtex4-2}
\usepackage{amsfonts,amsmath,amssymb,bm}
\usepackage{graphicx,graphics,float}
\usepackage{bm}
\usepackage{amssymb}
\usepackage{colordvi}
\usepackage{graphicx}
\usepackage{color}
\usepackage[colorlinks=true,linkcolor=blue,citecolor=blue,urlcolor=blue]{hyperref}%
\usepackage{hyperref}
\usepackage{comment}
\usepackage{harpoon}
\usepackage{braket}

\begin{document}


\title{Feedback-Enhanced Quantum Metrology and Clock Precision under Thermodynamic Uncertainty}

\author{Jincheng Lu}\email{jinchenglu@usts.edu.cn}
\affiliation{Key Laboratory of Intelligent Optoelectronic Devices and Chips of Jiangsu Higher Education Institutions, School of Physical Science and Technology, Suzhou University of Science and Technology, Suzhou, 215009, China}
\affiliation{Advanced Technology Research Institute of Taihu Photon Center, School of Physical Science and Technology, Suzhou University of Science and Technology, Suzhou, 215009, China}

%

\author{Chen Wang}\email{wangchen@zjnu.cn}
\affiliation{Department of Physics, Zhejiang Normal University, Jinhua, Zhejiang 321004, China}

\date{\today}

\begin{abstract}
Feedback can convert continuously monitored quantum jumps into a
thermodynamic resource. We formulate full counting statistics for open
quantum systems under  unital jump feedback by incorporating
the feedback maps into the tilted generator. The resulting trajectory
ensemble determines both current fluctuations and the Fisher information
of the measurement record. We show that feedback can enhance
reservoir-parameter estimation and clock precision without necessarily
changing average thermodynamic currents. This enhanced precision is not
bounded by reservoir entropy production alone. By embedding the reduced
dynamics in an enlarged measurement-feedback process, we derive a
feedback-modified thermodynamic uncertainty relation in which the
information entropy production of the feedback apparatus supplies the
missing cost. A charge-monitored double quantum dot illustrates the
framework: jump-conditioned feedback improves thermometry and
chemical-potential sensing, and stabilizes a quantum clock defined by
output-current ticks.
\end{abstract}

\maketitle
\emph{Introduction.}
Thermodynamic operation of quantum thermal machines is intrinsically accompanied with pronounced
fluctuations and the exchange of information~\cite{ParrondoNatPhys2015,horowitz20}, which may  not only determine the reliability conversion of
heat, energy, and work, but also improve the thermodynamic precision for sensing and metrology. 
In particular, current precision is generically constrained by thermodynamic uncertainty relations (TUR), which
connect low-order current fluctuations to entropy
production~\cite{TUR15,GingrichPRL}. 
Measurement precision is intrinsically trajectory-based, since exchanging events are inferred from time-resolved stochastic
records. 
Continuous quantum monitoring provides a direct access to these fluctuating trajectories~\cite{LandiPRXQuantum}.

In a monitored quantum device, the exchange of energy or particles with
the environment is resolved into a time-ordered quantum-jump record. This
record is simultaneously a thermodynamic readout and a metrological
resource, encoding transport fluctuations, temporal correlations, and the
sensitivity of the dynamics to reservoir parameters. When the measurement
outcomes are fed back to the system, the monitored jumps become active
control signals rather than passive observations~\cite{Wisemanbook}. Measurement-based
feedback can thus reshape the trajectory ensemble, stabilizing quantum
transport and modifying current fluctuations
\cite{BrandesPRL2010,EmaryGoughPRB2014}, waiting-time statistics
\cite{Stegmann}, and reservoir-parameter
sensitivity~\cite{GammelmarkPRL14,Radaelli26}. This interplay between
measurement, feedback, and trajectory fluctuations imperatively need a unified
description of metrological precision and thermodynamic cost.

Existing theories of feedback thermodynamics have explored fluctuation
relations, information flows, and entropy balances in a variety of
classical and quantum settings~\cite{UedaPRL08,TakahiroPRL10,HorowitzPRX14,YadaPRL22,PrechPRL24}. Yet the implications of
quantum-jump feedback for current precision and trajectory-level
parameter sensitivity remain less explored~\cite{TanPRXQuantum}. 
One knows that feedback acts directly on the
stochastic process that generates the counted events: it can leave the
jump weights entering the mean current unchanged, while altering the
steady state and the dynamical correlations that determine current noise.
Building on the feedback-transport perspective, we wonder how the controlled
trajectory ensemble carries Fisher information and clock precision, and what
information-thermodynamic cost restores the associated uncertainty bound.
Furthermore, capturing this distinction requires a full-counting-statistics
formulation, in which feedback operations, current cumulants, and the
Fisher information of the monitoring record are treated within a single
trajectory framework. Such a description is also needed to determine
whether feedback-enhanced sensing
precision and feedback-stabilized currents are sufficiently
bounded by the reservoir entropy production alone, or an
additional information-thermodynamic cost need
include~\cite{FunoPRE26,Funofeedback,RyotaroTUR}.

Here, we develop such a trajectory framework for continuously
monitored open quantum systems under instantaneous unital feedback. Each
detected jump triggers a completely positive trace-preserving operation,
which is incorporated into a tilted
Gorini-Kossakowski-Sudarshan-Lindblad
generator~\cite{lindblad,LandiRMP21}. The dominant eigenvalue of this
generator yields the long-time cumulants and large-deviation statistics
of time-integrated jump currents, while the same feedback-controlled
trajectory ensemble determines the Fisher information carried by the
monitoring record. This will establish a unified description of
feedback-controlled transport, reservoir-parameter sensing, and current
precision. Crucially, we show that the reservoir entropy production
is not the complete cost of
feedback-controlled precision. The feedback apparatus detects jumps,
processes the measurement record, applies conditional operations, and
resets its memory, thereby contributing an information-thermodynamic
cost 
to the enlarged
measurement-feedback process. 
The total precision cost
restores a TUR in which information
processing enters on an equal footing with thermal dissipation. 
We set $\hbar=k_B=1$ throughout.

\begin{figure}[t]
\centering
\includegraphics[width=\columnwidth]{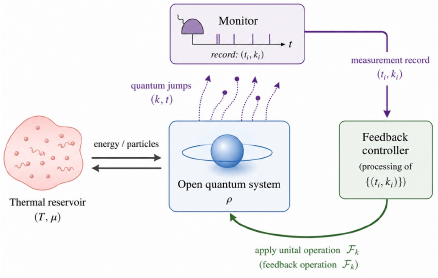}
\caption{Schematic of the quantum-jump feedback control setup.
An open quantum system (center) exchanges energy and/or particles
with a thermal reservoir, generating stochastic quantum jumps. A monitor continuously detects each jump and records its
type and occurrence time.  Conditioned on the detected jump type,
a feedback controller immediately applies a unital operation
$\mathcal{F}_k$ to the system.  The monitor and controller together form the feedback loop.}
\label{fig:setup}
\end{figure}

\emph{Full counting measurement under feedback.}
We consider an open quantum system weakly coupled to thermal reservoirs and continuously monitored via its dissipative jumps, as illustrated in Fig.~\ref{fig:setup}. 
The monitoring process generally resolves the system-reservoir exchanges into a time-ordered record of jump events. Given a quantum feedback protocol, each detected jump event is mapped immediately to modulate a control process within the evolving trajectory, which is embedded with the fluctuation.

Within the quantum-trajectory description, the monitored dynamics over an
infinitesimal time interval $dt$ is decomposed into jump and no-jump
branches. A jump of the channel $k\ge 1$ could be described by the Kraus operator
$M_k = L_k\sqrt{dt}$, while the no-jump evolution is generated by
$M_0 = \mathbb{I} - i H_{\rm eff}\,dt$, with the effective non-Hermitian
Hamiltonian $H_{\rm eff} = H - \frac{i}{2}\sum_k L_k^\dagger L_k$.
When jump-conditioned feedback is active, each detected jump of the channel $k$
is immediately followed by a completely positive, trace-preserving map
$\mathcal{F}_k$. The unconditional state then evolves as $\rho_{t+dt}
= M_0\rho_t M_0^\dagger + \sum_{k\ge 1} \mathcal{F}_k\bigl[ M_k\rho_t M_k^\dagger \bigr]$. Taking the continuum limit $dt\to 0$ yields the feedback master equation~\cite{lindblad,LandiRMP21}
\begin{eqnarray}
\dot\rho_t &&\equiv \mathcal{L}^{(\mathrm{fb})}[\rho_t]\label{eq:QME}\\
&&= i[\rho_t,H]+ \sum_{k\ge 1} \Bigl(
    \mathcal{F}_k[L_k\rho_t L_k^\dagger]
    - \frac12\{L_k^\dagger L_k,\rho_t\}
    \Bigr).\nonumber
\end{eqnarray}
Equation~\eqref{eq:QME} is the jump-conditioned version of the Markovian
feedback master equation developed within the quantum-trajectory and
continuous-measurement formalisms~\cite{carmichael1993,Wisemanbook},
and it is regarded as a standard starting point for feedback-controlled
quantum thermodynamics~\cite{RosalPRL2026,RosalPRA2026,RosalPRA262} and counting-statistics
measurement~\cite{LandiPRXQuantum}.

For each reservoir-induced dissipative channel $k$, there exists a reverse channel
$k^*$.
And the corresponding environmental entropy increments is known to satisfy
$\Delta s_{k^*} = -\Delta s_k$,
once bare jump operators fulfill local-detailed-balance relation~\cite{LandiPRXQuantum}. 
Specifically, for an electronic reservoir, this
entropy increment is fixed by the exchanged energy and particle number
according to the chosen current convention. In what follows, we focus on
unital feedback maps, $\mathcal{F}_k[\mathbb{I}] = \mathbb{I}$, which
include unitary feedback operations and nonselective 
measurements. In finite dimensions, such completely positive, trace-preserving unital maps do not decrease the von Neumann entropy. This property is useful for interpreting the feedback operation as an
information-thermodynamic control resource.
The consideration of energetic and
informational cost of the controller will be accounted for separately below.

Though Eq.~\eqref{eq:QME} describes the feedback-modified dynamics, the precision of transport, sensing, and clock ticks are actually encoded
in the fluctuations of trajectory observables. We therefore introduce the
full counting statistics, i.e., a representative approach to characterize fluctuations, of time-integrated jump currents. 
A monitored trajectory over a time interval $\tau$ is specified by a sequence of jump types $k_i$ and their occurrence times
$t_i$, denoted by $\Gamma = \{(t_1,k_1), \dots, (t_N,k_N)\}$~\cite{LandiPRXQuantum}.
For a current with increments $c_k$ satisfying $c_{k^*} = -c_k$, the
integrated current is $J(\Gamma) = \sum_{i=1}^N c_{k_i}$ and its
moment-generating function is $Z(\chi) = \langle \exp{(i\chi J)}\rangle$.
In the continuous-time limit, $Z(\chi)$ can be generated from the tilted
feedback master equation~\cite{LandiRMP21}
$\dot\rho_{t,\chi}
= \mathcal{L}_\chi^{(\mathrm{fb})}[\rho_{t,\chi}]$, i.e.,
\begin{equation}
\dot\rho_{t,\chi}
= \mathcal{L}^{(\mathrm{fb})}[\rho_{t,\chi}]
  + \sum_{k\ge 1} \bigl( e^{i\chi c_k} - 1 \bigr)\,
    \mathcal{F}_k\bigl[ L_k \rho_{t,\chi} L_k^\dagger \bigr],
\label{eq:Lchi}
\end{equation}
where $\mathcal{L}^{(\mathrm{fb})}[\cdot]$ is the feedback Liouvillian operator. The generating function is reexpressed as
$Z(\chi) = \operatorname{tr}[\rho_{\tau,\chi}]$, and in the long-time
limit $Z(\chi) \asymp e^{\tau\lambda_{\mathrm{fb}}(\chi)}$, where
$\lambda_{\mathrm{fb}}(\chi)$ is the eigenvalue of
$\mathcal{L}_\chi^{(\mathrm{fb})}$ with the largest real part.

As is known that the long-time current cumulants follow from the scaled cumulant
generating function~\cite{fluctuationRMP,LandiPRXQuantum}. 
Specifically, the steady-state current and the
zero-frequency noise
are given by
$\langle  J\rangle_{\mathrm{fb}}
= \partial_{i\chi}\lambda_{\mathrm{fb}}(\chi)\big|_{\chi=0}$,
and
$\langle\!\langle J^2\rangle\!\rangle_{\mathrm{fb}}
= \partial_{i\chi}^{2}\lambda_{\mathrm{fb}}(\chi)\big|_{\chi=0}$, respectively.
The first derivative can be written equivalently as the jump-rate
expression
$\langle J\rangle_{\mathrm{fb}}
= \sum_{k\ge 1} c_k
\operatorname{tr}\bigl[ L_k \rho_{\mathrm{ss}}^{(\mathrm{fb})} L_k^\dagger \bigr]$,
where $\mathcal{L}^{(\mathrm{fb})}[\rho_{\mathrm{ss}}^{(\mathrm{fb})}]=0$.
We note the feedback map does not appear explicitly in this jump weight because
every $\mathcal{F}_k$ is trace preserving. Nevertheless, feedback
changes the steady state $\rho_{\mathrm{ss}}^{(\mathrm{fb})}$ and, more
importantly, enters the higher cumulants through the tilted generator
$\mathcal{L}_\chi^{(\mathrm{fb})}$. 
Thus, feedback provides a direct
route to reshaping current fluctuations while preserving a compact
trajectory-level full-counting structure.

\emph{Parameter estimation from feedback-controlled quantum-jump records.}
The full-counting-statistics framework characterizes the cumulants of
time-integrated currents in the monitored open system. The same
quantum-jump record can also be used to estimate unknown parameters of
the system or its environment. We consider a parameter \(\theta\) entering
the Hamiltonian \(H(\theta)\) and/or the jump operators \(L_k(\theta)\).
In the applications below, \(\theta\) represents a reservoir property,
such as a temperature or chemical potential, whose influence is encoded
in the dissipative jump process.

We consider a prescribed feedback strategy that processes the detected
jump record independently of the reservoir parameter \(\theta\), yielding
\(\partial_\theta \mathcal F_k=0\). As a standard consequence of
classical estimation theory~\cite{Rao1945,Cramer1946}, the Fisher
information of the feedback-controlled trajectory ensemble sets the
Cram\'er--Rao bound for any unbiased estimator \(\hat\theta\) constructed
from the monitored record,
\({\rm Var}(\hat\theta)\ge 1/\mathcal F_{\rm fb}(\theta)\).
The estimation precision is therefore governed by the parameter
sensitivity of the full monitored trajectory ensemble, not by a single
current cumulant alone.

For a sufficiently long monitoring time \(\tau\), the Fisher information
of the feedback-controlled quantum-jump record takes the asymptotic
form~\cite{GammelmarkPRL14,Radaelli26}; a derivation for the present
jump-conditioned feedback dynamics is given in Sec.~S2 of the
Supplemental Material~\cite{SM},
\begin{equation}
\begin{aligned}
&\frac{\mathcal F_{\rm fb}(\theta)}{4\tau}
=
\sum_k
\langle
(\partial_\theta L_k)^\dagger(\partial_\theta L_k)
\rangle_{\rm ss}^{({\rm fb})}\\
&\quad
-
\langle\!\langle\mathbb I|
\mathcal L_L^{({\rm fb})}\mathcal L^{({\rm fb})+}
\mathcal L_R^{({\rm fb})}
|\rho_{\rm ss}^{({\rm fb})}\rangle\!\rangle -
\langle\!\langle\mathbb I|
\mathcal L_R^{({\rm fb})}\mathcal L^{({\rm fb})+}
\mathcal L_L^{({\rm fb})}
|\rho_{\rm ss}^{({\rm fb})}\rangle\!\rangle .
\end{aligned}
\label{eq:FI_fb}
\end{equation}
Here
\(\langle A\rangle_{\rm ss}^{({\rm fb})}
=
{\rm Tr}[A\rho_{\rm ss}^{({\rm fb})}]\),
\(|\rho_{\rm ss}^{({\rm fb})}\rangle\!\rangle\) is the vectorized
steady state, \(\langle\!\langle\mathbb I|\) is the trace functional, and
\(\mathcal L^{({\rm fb})+}\) is the Drazin inverse of the feedback
Liouvillian. The superoperators
\(\mathcal L_{L,R}^{({\rm fb})}\) collect parameter derivatives acting
on the no-jump evolution and on the jump amplitudes, with the feedback
maps inserted in the jump branches. Their explicit matrix forms are given in Sec.~S2 of the Supplemental Material~\cite{SM}.

Equation~\eqref{eq:FI_fb} is a central theoretical result of this work.
It extends the trajectory Fisher-information description of monitored
quantum jumps~\cite{GammelmarkPRL14,Radaelli26} to
feedback-controlled trajectory ensembles. Feedback enters this expression
in two ways: it changes the stationary state
\(\rho_{\rm ss}^{({\rm fb})}\), and it appears explicitly in the
propagation of parameter perturbations through
\(\mathcal L_{L,R}^{({\rm fb})}\). Thus the measurement record is not
only passively read out, but actively reshaped by the feedback loop
before its parameter sensitivity is evaluated. When the feedback maps are
replaced by the identity channel, \(\mathcal F_k[\rho]=\rho\), the
standard feedback-free result is recovered.


\begin{figure}[htb]
\begin{center}
\centering\includegraphics[width=8.5cm]{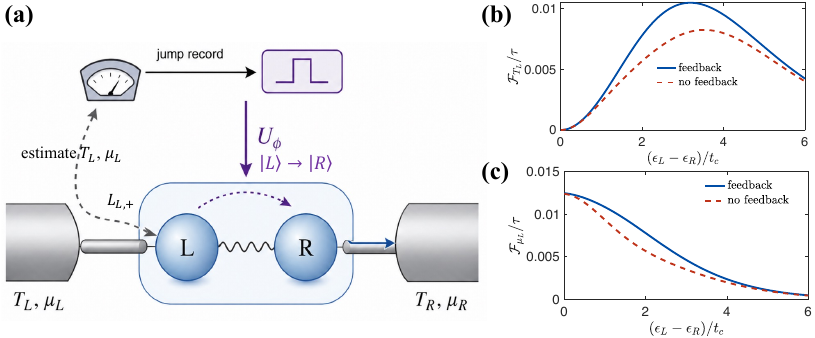}
\caption{Feedback-enhanced thermometry with a charge-monitored DQD.
(a) Schematic of the setup. The DQD is coupled to two electronic reservoirs
at temperatures $T_L$, $T_R$ and chemical potentials $\mu_L$, $\mu_R$.
A left-injection jump $L_{L,+}$ is continuously monitored.
Conditioned on its detection, a fast unitary feedback pulse $U_\phi$
rotates the post-jump state from the left dot toward the right dot,
reshaping the subsequent waiting-time and jump statistics.
The resulting feedback-controlled jump record is used to estimate
$T_L$. (b) Fisher information rate $\mathcal{F}_{T_L}/\tau$ for estimating
the left-reservoir temperature $T_L$ versus the DQD detuning
$\epsilon_L-\epsilon_R$, with $\mu_L=\mu_R=0$ and the average dot
energy fixed at $(\epsilon_L+\epsilon_R)/2=0$.
(c) Fisher information rate $\mathcal{F}_{\mu_L}/\tau$ for estimating
the left chemical potential $\mu_L$ under the same conditions, with
$T_L=T_R=1$. Solid (dashed) curves correspond to the feedback-controlled
(no-feedback) case. Parameters: $t_c=1.0$, $\Gamma_L=0.2$, $\Gamma_R=0.6$,
$T_L=T_R=1$, $\mu_L=\mu_R=0$, and $\phi=\pi/2$.}
\label{fig:figure2}
\end{center}
\end{figure}

We exemplify this mechanism with a minimal charge-monitored double
quantum dots in Fig.~\ref{fig:figure2}(a). Such device provides
a practical platform for jump-based feedback
~\cite{FujisawaScience2006,GustavssonPRL2006}.
Meanwhile, charge-detected DQD
feedback and information-to-work conversion have been widely discussed
theoretically~\cite{Hussein2014,NakajimaPRApplied2021,
AnnbyAndersson2020,TrigalSanchezPRB2026}. 
The DQD, denoting \(|0\rangle\), \(|L\rangle\), and \(|R\rangle\), is coupled to two
electronic reservoirs, and left-injection events are continuously
monitored. Whenever an \(L_{L,+}\) jump event is detected, a fast conditional
pulse \(U_\phi\) rotates the post-jump state from the left dot toward the
right dot. 
This pulse may be implemented as a fast gate-controlled charge
rotation, such as a Landau-Zener-type passage or a resonant charge-qubit
pulse.
The explicit DQD Hamiltonian, jump operators, feedback map, and tilted
generator used in the numerical calculations are given in Sec.~S4 of the
Supplemental Material, while the derivation of the feedback-modified
trajectory Fisher information is given in Sec.~S2~\cite{SM}.

Figure~\ref{fig:figure2}(b) shows the Fisher information rate for
estimating the left-reservoir temperature $T_L$ as a function of the DQD detuning. 
It is intriguing to find that feedback enhances the temperature sensitivity at an optimal energy detuning window, i.e., $\mathcal{F}_{T_L}/\tau$ exceeding the no-feedback reference for
$1 \lesssim (\epsilon_L-\epsilon_R)/t_c \lesssim 6$.
For larger detunings the two curves nearly overlap, indicating the vanish of the feedback advantage 
Figure~\ref{fig:figure2}(c) shows the Fisher information rate for estimating the left chemical potential \(\mu_L\). The feedback-controlled trajectory ensemble generally retains a high sensitivity to \(\mu_L\).
As the DQD detuning increases, the feedback curve becomes robust comparing with the no-feedback reference, which rapidly loses parameter sensitivity. 
Together, these two panels demonstrate that the feedback pulse affects
distinct reservoir parameters in corresponding routes. For $T_L$, it enhances
the Fisher information most strongly near the optimal detuning. 
While for
$\mu_L$, it extends the useful parameter-estimation regime by slowing
the loss of sensitivity with increasing level mismatch. This positively contributes to
the interpretation that jump-conditioned feedback reshapes the monitored
trajectory ensemble and amplifies its parameter sensitivity, rather than
merely increasing the overall jump rate.

\emph{Feedback-modified TUR.} Under jump-conditioned feedback, each detected jump operation
\(\mathcal J_k[\rho]=L_k\rho L_k^\dagger\) is followed by a conditional
map \(\mathcal F_k\). The feedback-transformed forward and backward
channels are therefore generally not related by the same
detailed-balance conjugation as the bare reservoir jumps.
Consequently, the reduced tilted generator need not obey the conventional
fluctuation symmetry that holds in feedback-free dynamics under local
detailed balance~\cite{fluctuationRMP,LebowitzSpohn1999}.
This is not a thermodynamic loophole. It means that the
entropy production of the reservoirs alone is not the complete cost of
the reduced feedback-controlled dynamics.

To restore the thermodynamic bookkeeping, we embed the
reduced dynamics in an enlarged measurement-feedback process.
Irreversibility in this enlarged process has two physically distinct
origins: reservoir-induced transitions and controller-induced
transitions. A reservoir-induced jump channel \(k\) occurs
with stationary rate
\(R_k^{({\rm fb})}
=
{\rm Tr}[L_k\rho_{\rm ss}^{({\rm fb})}L_k^\dagger]\)
and changes the reservoir entropy by \(\Delta s_k\), giving
\(\dot\Sigma_{\rm th}
=
\sum_{k\in{\rm th}}R_k^{({\rm fb})}\Delta s_k\).
The controller-induced contribution is governed by a different rate,
\(R_{\rm trig}^{({\rm fb})}\), namely the total stationary occurrence
rate of the jump channels that activate the feedback loop.
The effective entropy production rate entering the
feedback-modified TUR is
\begin{equation}
\dot\Sigma_{\rm eff}
=
\dot\Sigma_{\rm th}
+
\dot\Sigma_{\rm fb}^{\rm info}.
\label{eq:Sigma_eff}
\end{equation}
Here
\(\dot\Sigma_{\rm fb}^{\rm info}
=
R_{\rm trig}^{({\rm fb})}\sigma_{\rm fb}\)
is the information entropy production rate of the feedback apparatus,
and \(\sigma_{\rm fb}\) is the entropy cost of one complete
detection--control--reset cycle. The minimal-controller
decomposition of \(\sigma_{\rm fb}\), and the entropic-perturbation
derivation of the bound below, are given in Sec.~S3 of the Supplemental
Material~\cite{SM}. In the DQD application, the only
feedback-triggering channel is the left-injection jump \(L_{L,+}\), so
that \(R_{\rm trig}^{({\rm fb})}=R_{L,+}^{({\rm fb})}\).
The value of \(\sigma_{\rm fb}\) used below is fixed by
this controller model and is not adjusted to enforce the inequality.

Applying the Cram\'er--Rao construction to an entropic perturbation of
the enlarged trajectory ensemble yields
\begin{equation}
\frac{
\langle\!\langle J^2\rangle\!\rangle_{\rm fb}
}{
\langle\dot J\rangle_{\rm fb}^{2}
}
\ge
\frac{2}{\dot\Sigma_{\rm eff}} .
\label{eq:feedback_modified_TUR}
\end{equation}
Equation~\eqref{eq:feedback_modified_TUR} shows that
feedback-enhanced current precision is bounded by the entropy production
of the enlarged measurement-feedback process, not by the reservoir
entropy production of the reduced system alone.

\begin{figure}[htb]
\begin{center}
\centering\includegraphics[width=8.5cm]{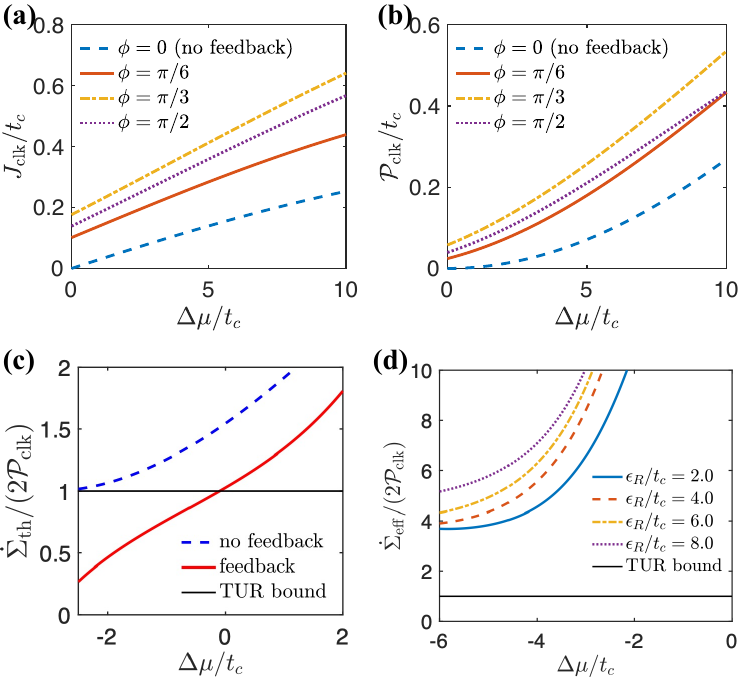}
\caption{
Feedback-stabilized DQD quantum clock and its thermodynamic cost.
Right-reservoir output events are interpreted as clock ticks, with
$J_{\rm clk}$ the net tick rate and $S_{\rm clk}$ the corresponding
zero-frequency noise.
(a) Tick rate $J_{\rm clk}/t_c$ versus the chemical-potential bias
$\Delta\mu=\mu_L-\mu_R$ for several feedback angles $\phi$.
(b) Clock precision $\mathcal{P}_{\rm clk}/t_c$ for the same set of $\phi$.
(c) Reservoir-only TUR ratio
$\dot\Sigma_{\rm th}/(2\mathcal{P}_{\rm clk})$ versus
$\Delta\mu/t_c$. Values below the horizontal line indicate that the
thermal entropy production alone is insufficient to bound the
feedback-stabilized clock precision.
(d) Effective TUR ratio
$\dot\Sigma_{\rm eff}/(2\mathcal{P}_{\rm clk})$.
The feedback entropy production is
$\dot\Sigma_{\rm fb}^{\rm info}
= R_{L,+}^{({\rm fb})}\sigma_{\rm fb}$, where
$R_{L,+}^{({\rm fb})}$ is the rate of the feedback-triggering
left-injection jumps and $\sigma_{\rm fb}$ is the entropy cost per
feedback cycle, taken as
$\sigma_{\rm fb}
= \ln 2 + \beta_c[(\epsilon_R-\epsilon_L)\sin^2\phi]_+$,
which includes a Landauer-scale binary-record reset cost and the
energetic cost of the conditional pulse.
Unless stated otherwise, $(\mu_L+\mu_R)/2=0$, $T_L=T_R=1$, and
$\hbar=k_B=1$. In (a,b): $\epsilon_L=\epsilon_R=0$, $t_c=0.2$, $\Gamma_L=0.4$, $\Gamma_R=1.0$.}
\label{fig:figure3}
\end{center}
\end{figure}

\emph{Feedback-stabilized quantum clock.} The same feedback protocol can also serve as a timing stabilizer, effectively turning the device into a minimal quantum clock~\cite{ErkerPRX17,SchwarzhansPRX2021}. The feedback protocol is unchanged: a detected left-injection jump $L_{L,+}$ triggers the unitary pulse $U_\phi$, which rotates the post-jump state as $U_\phi\ket{L} = \cos\phi\,\ket{L} - i\sin\phi\,\ket{R}$. For $\phi\simeq\pi/2$, the electron is transferred close to $\ket{R}$,
from which it can escape into the right reservoir.

A clock tick can be defined as a right-reservoir emission event $L_{R,-}$~\cite{PrechPRX25}.
The tick statistics are obtained from the tilted feedback Liouvillian
with a counting field conjugate to the tick number; the dominant
eigenvalue yields the tick rate $\dot N_{\rm clk}$ and the zero-frequency
noise $S_{\rm clk}$. The clock precision rate is
$\mathcal{P}_{\rm clk}
= \frac{\dot N_{\rm clk}^{2}}{S_{\rm clk}}$,
so that a smaller Fano factor $S_{\rm clk}/\dot N_{\rm clk}$ corresponds
to more regular ticks. The feedback-modified thermodynamic uncertainty
relation then bounds this precision as
\begin{equation}
\mathcal{P}_{\rm clk}
\le \frac{\dot\Sigma_{\mathrm{eff}}}{2}.
\label{eq:clock_precision_bound}
\end{equation}

Figures~\ref{fig:figure3}(a) and \ref{fig:figure3}(b) shows how the feedback angle controls the clock performance. The tick rate $J_{\rm clk}$ is enhanced once the left-injection jump is followed by a finite feedback rotation, but the enhancement is not monotonic in $\phi$. In the present parameter regime, $J_{\rm clk}$ reaches its maximum near $\phi=\pi/3$, while a stronger rotation to $\phi=\pi/2$ slightly reduces the output tick rate. The
precision rate $\mathcal{P}_{\rm clk}=J_{\rm clk}^2/S_{\rm clk}$ shows
the same tendency, also peaking near $\phi=\pi/3$. This nonmonotonic
behavior indicates that the best clock performance is not achieved by a maximal pulse, but by an optimal balance between state preparation,
coherent tunneling, and escape into the right reservoir. Figures~\ref{fig:figure3}(c) and \ref{fig:figure3}(d) test the thermodynamic cost of this
feedback-stabilized precision. Figure~\ref{fig:figure3}(c) shows the
reservoir-only TUR ratio
$\dot\Sigma_{\rm th}/(2\mathcal{P}_{\rm clk})$. Without feedback, this ratio remains above unity, consistent with the conventional TUR. Once feedback is switched on, the same ratio drops below unity. This does not imply that feedback escapes thermodynamic constraints; rather, it shows that the reservoir entropy production $\dot\Sigma_{\rm th}$ alone is no
longer the complete cost of producing a precise clock signal. Figure~\ref{fig:figure3}(d) verifies the feedback-modified bound. There the effective ratio $\dot\Sigma_{\rm eff}/(2\mathcal{P}_{\rm clk})$,
which includes the information-processing cost
$\dot\Sigma_{\rm fb}^{\rm info}$ through
$\dot\Sigma_{\rm eff} = \dot\Sigma_{\rm th} + \dot\Sigma_{\rm fb}^{\rm info}$,
remains above unity for all feedback angles considered. The comparison
between Figs.~\ref{fig:figure3} (c) and \ref{fig:figure3}(d) demonstrates that the improved clock regularity is not free: it is supported by the information processing and conditional control performed by the feedback loop.

\emph{Conclusion.} We have developed a feedback-controlled quantum-jump trajectory framework that incorporates full counting measurement, trajectory Fisher
information, TUR, and clock precision within a unified description. The
theory demonstrates that jump-conditioned feedback can generally enhance
reservoir-parameter sensing and stabilize fluctuating output events as
regular clock ticks by reshaping the monitored trajectory ensemble.
This precision gain is not bounded by the reservoir entropy production
alone. By embedding the dynamics in an enlarged measurement-feedback
process, we identified that the information entropy production of the
feedback apparatus as the missing thermodynamic cost and correctly restored the
thermodynamic uncertainty relation. These results establish
feedback-mediated information flow as a thermodynamic resource for
quantum metrology, clock stabilization, and the design of
precision-cost trade-offs in monitored nonequilibrium quantum devices.

\emph{Acknowledgments.} We are deeply indebted to R. S\'anchez for generously reading an early version of this manuscript and for sharing detailed comments that considerably sharpened both the physical picture and the presentation of this work. We acknowledge support from the National Natural Science Foundation of China (Grant No. 12305050 and No. 12674313), the Natural Science Foundation of Jiangsu Higher Education Institutions of China (Grant No. 23KJB140017), and the Zhejiang Provincial Natural Science Foundation of China under Grant No. LZ25A050001.

\bibliography{Reffeedback}

\clearpage  

\onecolumngrid  
\section*{Supplemental Material: Feedback-Enhanced Quantum Metrology and Clock Precision under Thermodynamic Uncertainty}

\tableofcontents

\setcounter{section}{0}
\renewcommand{\thesection}{S\arabic{section}}
\renewcommand{\theequation}{S\arabic{equation}}  
\setcounter{equation}{0}  

\section{Derivation of the tilted feedback quantum master equation}
\label{sec:SM_tilted}

We provide a step-by-step derivation of Eq.~\eqref{eq:Lchi} of the main
text.


Divide the total evolution time $\tau$ into $N$ steps of length $dt$
with $\tau = N dt$.  In each step the system evolves according to the
Kraus operators
\begin{equation}
M_0 = \mathbb{I} - \bigl(iH + \tfrac12\sum_{k\ge 1} L_k^\dagger L_k\bigr)dt,
\qquad
M_k = L_k\sqrt{dt}\ \ (k\ge 1).
\label{eq:Mk}
\end{equation}
If a jump of type $k$ occurs, the conditional state is updated by
$M_k$ and subsequently by the feedback map $\mathcal{F}_k$.
If no jump occurs, only $M_0$ is applied.

A trajectory over the entire interval is specified by a sequence
$(k_1, k_2, \dots, k_N)$, where $k_i = 0$ denotes no jump and
$k_i \ge 1$ denotes a jump of type $k_i$ in the $i$th step.
The probability of a given trajectory is
\begin{equation}
p(\Gamma) = \operatorname{tr}\!\Bigl[
\mathcal{F}_{k_N}\!\bigl( M_{k_N} \cdots
\mathcal{F}_{k_1}\!\bigl( M_{k_1} \rho_0 M_{k_1}^\dagger \bigr)
\cdots M_{k_N}^\dagger \bigr)
\Bigr],
\label{eq:p_Gamma}
\end{equation}
and the integrated current reads
$J(\Gamma) = \sum_{i=1}^N c_{k_i}$ with $c_0 \equiv 0$.


The moment-generating function is
$Z(\chi) = \sum_\Gamma e^{i\chi J(\Gamma)} p(\Gamma)$.
To absorb the exponential weight into the dynamics, define the tilted
Kraus operators
\begin{equation}
\tilde M_0 = M_0, \qquad
\tilde M_k = e^{i\chi c_k/2} M_k\ \ (k\ge 1).
\label{eq:tilde_M}
\end{equation}
The generating function then takes the compact form
\begin{equation}
Z(\chi) = \operatorname{tr}\!\Bigl[
\sum_{k_N}\!\cdots\!\sum_{k_1}
\mathcal{F}_{k_N}\!\bigl( \tilde M_{k_N} \cdots
\mathcal{F}_{k_1}\!\bigl( \tilde M_{k_1} \rho_0
\tilde M_{k_1}^\dagger \bigr) \cdots
\tilde M_{k_N}^\dagger \bigr)
\Bigr].
\label{eq:Z_disc}
\end{equation}
Introducing the tilted density operator after $n$ steps via the
recurrence
\begin{equation}
\rho_{n dt,\chi}
= \sum_{k\ge 0} \mathcal{F}_k\!\bigl(
   \tilde M_k\, \rho_{(n-1)dt,\chi}\, \tilde M_k^\dagger \bigr),
\qquad \rho_{0,\chi} = \rho_0,
\label{eq:recurrence}
\end{equation}
we obtain $Z(\chi) = \operatorname{tr}[\rho_{\tau,\chi}]$.


Expanding $\tilde M_0$ and $\tilde M_k$ to first order in $dt$ yields
\begin{subequations}
\begin{align}
\tilde M_0 \rho \tilde M_0^\dagger
&= \rho + \Bigl( -i[H,\rho]
   - \frac12\sum_{k\ge 1}\{L_k^\dagger L_k,\rho\} \Bigr) dt
   + \mathcal{O}(dt^2), \label{eq:exp_M0} \\
\tilde M_k \rho \tilde M_k^\dagger
&= e^{i\chi c_k} L_k \rho L_k^\dagger dt
   + \mathcal{O}(dt^2) \quad (k\ge 1). \label{eq:exp_Mk}
\end{align}
\end{subequations}
Substituting Eqs.~\eqref{eq:exp_M0} and~\eqref{eq:exp_Mk} into the
recurrence~\eqref{eq:recurrence} and keeping the terms of order $dt$,
we obtain in the limit $dt\to 0$
\begin{equation}
\frac{d\rho_{t,\chi}}{dt}
= -i[H,\rho_{t,\chi}]
  + \sum_{k\ge 1} \Bigl(
     e^{i\chi c_k} \mathcal{F}_k[L_k \rho_{t,\chi} L_k^\dagger]
     - \frac12\{L_k^\dagger L_k, \rho_{t,\chi}\}
     \Bigr).
\label{eq:tilted_master}
\end{equation}
This is the tilted feedback master equation.  It can be rewritten in
terms of the tilted superoperator as
$\dot{\rho}_{t,\chi} = \mathcal{L}_\chi^{(\mathrm{fb})}[\rho_{t,\chi}]$,
with
\begin{equation}
\mathcal{L}_\chi^{(\mathrm{fb})}(\circ)
= \mathcal{L}^{(\mathrm{fb})}(\circ)
+ \sum_{k\ge 1} (e^{i\chi c_k} - 1)\,
  \mathcal{F}_k[L_k \circ L_k^\dagger],
\label{eq:Lchi_SM}
\end{equation}
which is Eq.~\eqref{eq:Lchi} of the main text.


The average current is $\langle J\rangle
= -i\partial_\chi Z(\chi)|_{\chi=0}$.
Using $\partial_\chi \mathcal{L}_\chi^{(\mathrm{fb})}|_{\chi=0}
= \sum_k i c_k \mathcal{F}_k[L_k \cdot L_k^\dagger]$ and the trace
preservation of $\exp[t\mathcal{L}^{(\mathrm{fb})}]$, one finds
\begin{align}
\langle J \rangle
&= \int_0^\tau dt\,
   (-i\partial_\chi) \operatorname{tr}\!\bigl[
   \mathcal{L}_\chi^{(\mathrm{fb})}(\rho_{t,\chi}) \bigr] \big|_{\chi=0}
   \nonumber \\
&= \int_0^\tau dt \sum_{k\ge 1} c_k
   \operatorname{tr}\!\bigl(
   \mathcal{F}_k[L_k \rho_t L_k^\dagger] \bigr) \nonumber \\
&= \int_0^\tau dt \sum_{k\ge 1} c_k
   \operatorname{tr}\!\bigl( L_k \rho_t L_k^\dagger \bigr).
\label{eq:avg_J_SM}
\end{align}
The last equality holds because each $\mathcal{F}_k$ is trace-preserving,
which allows us to drop the explicit feedback dependence at the level
of the average current.

The variance of the time-integrated current $J$ is obtained from the
second derivative of the generating function,
$\langle\!\langle J^2 \rangle\!\rangle
= -\partial_\chi^2 \ln Z(\chi)\big|_{\chi=0}$.
A direct differentiation of $Z(\chi) = \operatorname{tr}[\rho_{\tau,\chi}]$
gives
\begin{equation}
\langle\!\langle J^2 \rangle\!\rangle
= \langle J^2 \rangle - \langle J \rangle^2
= -\partial_\chi^2 \operatorname{tr}[\rho_{\tau,\chi}]\big|_{\chi=0}
  + \bigl( \partial_\chi \operatorname{tr}[\rho_{\tau,\chi}]\big|_{\chi=0}
  \bigr)^2.
\label{eq:var_def}
\end{equation}

To evaluate the derivatives, we use the formal solution
$\rho_{\tau,\chi} = e^{\tau \mathcal{L}_\chi^{(\mathrm{fb})}} \rho_0$.
Expanding the exponential in a Dyson series around $\chi=0$ yields
\begin{equation}
\begin{aligned}
\rho_{\tau,\chi}
&= \rho_\tau
   + \int_0^\tau dt\,
     e^{(\tau-t)\mathcal{L}^{(\mathrm{fb})}}\,
     \partial_\chi\mathcal{L}_\chi^{(\mathrm{fb})}\big|_{\chi=0}\,
     e^{t\mathcal{L}^{(\mathrm{fb})}} \rho_0 \;\; \chi \\
&\quad + \frac12 \int_0^\tau dt\,
     e^{(\tau-t)\mathcal{L}^{(\mathrm{fb})}}\,
     \partial_\chi^2\mathcal{L}_\chi^{(\mathrm{fb})}\big|_{\chi=0}\,
     e^{t\mathcal{L}^{(\mathrm{fb})}} \rho_0 \;\; \chi^2 \\
&\quad + \int_0^\tau dt \int_0^t ds\,
     e^{(\tau-t)\mathcal{L}^{(\mathrm{fb})}}\,
     \partial_\chi\mathcal{L}_\chi^{(\mathrm{fb})}\big|_{\chi=0}\,
     e^{(t-s)\mathcal{L}^{(\mathrm{fb})}}\,
     \partial_\chi\mathcal{L}_\chi^{(\mathrm{fb})}\big|_{\chi=0}\,
     e^{s\mathcal{L}^{(\mathrm{fb})}} \rho_0 \;\; \chi^2 \\
&\quad + \mathcal{O}(\chi^3),
\end{aligned}
\label{eq:Dyson}
\end{equation}
where $\rho_\tau = e^{\tau\mathcal{L}^{(\mathrm{fb})}}\rho_0$ is the
unperturbed state at time $\tau$.

Taking the trace and using $\operatorname{tr}[\mathcal{L}^{(\mathrm{fb})}(\circ)]=0$,
the terms linear in $\chi$ vanish at $\chi=0$ after subtracting
$\langle J\rangle^2$.  The quadratic term gives the variance.
In the long-time limit $\tau\to\infty$, the dominant contribution comes
from the double integral, leading to the steady-state variance rate
\begin{equation}
\langle\!\langle J^2 \rangle\!\rangle
\equiv \lim_{\tau\to\infty} \frac{\mathrm{Var}(J)}{\tau}
= \sum_{k\ge 1} c_k^2
   \operatorname{tr}\!\bigl( L_k \bar\rho L_k^\dagger \bigr)
   - 2 \sum_{k,\ell\ge 1} c_k c_\ell\,
   \operatorname{tr}\!\Bigl[
   \mathcal{F}_k[L_k \bar\rho L_k^\dagger]\,
   \mathcal{L}^{(\mathrm{fb})+}\,
   \mathcal{F}_\ell[L_\ell \bar\rho L_\ell^\dagger]
   \Bigr],
\label{eq:var_ss}
\end{equation}
where $\bar\rho$ is the steady-state density matrix satisfying
$\mathcal{L}^{(\mathrm{fb})}\bar\rho = 0$, and
$\mathcal{L}^{(\mathrm{fb})+}$ is the Drazin inverse of the feedback
Liouvillian.  The first term on the right-hand side of
Eq.~\eqref{eq:var_ss} is the Poissonian contribution from uncorrelated
jumps.  The second term, involving the Drazin inverse, encodes the
correlations between jumps induced by the interplay of the Hamiltonian
evolution, the dissipative dynamics, and the feedback operations.

Equation~\eqref{eq:var_ss} can be derived more directly from the
eigenvalue perturbation of the tilted Liouvillian.
Let $\lambda_{\mathrm{max}}(\chi)$ be the eigenvalue of
$\mathcal{L}_\chi^{(\mathrm{fb})}$ with the largest real part,
so that $\ln Z(\chi) \simeq \tau \lambda_{\mathrm{max}}(\chi)$
for large $\tau$.  Expanding
$\lambda_{\mathrm{max}}(\chi) = \lambda^{(0)} + i\chi\lambda^{(1)}
- \frac12 \chi^2 \lambda^{(2)} + \mathcal{O}(\chi^3)$,
with $\lambda^{(0)} = 0$ (trace preservation),
$\lambda^{(1)} = \langle \dot J \rangle$,
and $\lambda^{(2)} = \langle\!\langle J^2 \rangle\!\rangle$,
the standard Rayleigh-Schr\"odinger perturbation theory for non-Hermitian
operators gives the following
\begin{subequations}
\begin{align}
\lambda^{(1)}
&= \sum_{k\ge 1} c_k
   \langle\!\langle \mathbb{I} |
   \mathcal{F}_k[L_k \bar\rho L_k^\dagger]
   \rangle\!\rangle, \label{eq:lambda1} \\
\lambda^{(2)}
&= \sum_{k\ge 1} c_k^2
   \langle\!\langle \mathbb{I} |
   \mathcal{F}_k[L_k \bar\rho L_k^\dagger]
   \rangle\!\rangle
   - 2 \sum_{k,\ell\ge 1} c_k c_\ell\,
   \langle\!\langle \mathbb{I} |
   \mathcal{F}_k[L_k \bar\rho L_k^\dagger]\,
   \mathcal{L}^{(\mathrm{fb})+}\,
   \mathcal{F}_\ell[L_\ell \bar\rho L_\ell^\dagger]
   \rangle\!\rangle.
\label{eq:lambda2}
\end{align}
\end{subequations}
Since each $\mathcal{F}_k$ is trace-preserving,
$\langle\!\langle \mathbb{I} | \mathcal{F}_k[L_k \bar\rho L_k^\dagger]
\rangle\!\rangle = \operatorname{tr}(L_k \bar\rho L_k^\dagger)$,
recovering Eq.~\eqref{eq:avg_J_SM} for the average current.
For the variance, Eq.~\eqref{eq:lambda2} coincides with
Eq.~\eqref{eq:var_ss}, and the feedback maps $\mathcal{F}_k$ remain
explicitly present in the second term.  This is in stark contrast to
the average current, where trace preservation removes the $\mathcal{F}_k$
dependence.  The persistence of $\mathcal{F}_k$ in the variance is the
mathematical origin of the feedback's ability to modify fluctuations
without altering the mean.

\section{Trajectory Fisher information under jump-conditioned feedback}
\label{sec:SM_FI_feedback}

We derive the Fisher information of continuously monitored quantum-jump
records in the presence of jump-conditioned feedback. The quantity
considered here is the classical Fisher information of the observed
trajectory ensemble. It should be distinguished from the quantum Fisher
information optimized over all possible measurements.

\subsection{Feedback-free trajectory Fisher information}

We first recall the standard result for a continuously monitored
Markovian open system. The reduced state obeys
\[
\dot\rho
=
\mathcal L[\rho]
=
-i[H,\rho]
+
\sum_k
\left(
L_k\rho L_k^\dagger
-
\frac12\{L_k^\dagger L_k,\rho\}
\right),
\]
where the Hamiltonian and jump operators may depend on an unknown
parameter \(\theta\). For a long monitoring time \(\tau\), the Fisher
information of the quantum-jump record is
\begin{equation}
\begin{aligned}
\frac{\mathcal F(\theta)}{4\tau}
&=
\sum_k
\left\langle
(\partial_\theta L_k)^\dagger
(\partial_\theta L_k)
\right\rangle_{\rm ss}
\\
&\quad
-
\langle\!\langle\mathbb I|
\mathcal L_L
\mathcal L^+
\mathcal L_R
|\rho_{\rm ss}\rangle\!\rangle
-
\langle\!\langle\mathbb I|
\mathcal L_R
\mathcal L^+
\mathcal L_L
|\rho_{\rm ss}\rangle\!\rangle .
\end{aligned}
\label{eq:SM_FI_no_fb}
\end{equation}
Here
\(\langle A\rangle_{\rm ss}={\rm Tr}(A\rho_{\rm ss})\),
\(|\rho_{\rm ss}\rangle\!\rangle\) is the vectorized steady state,
\(\langle\!\langle\mathbb I|\) is the trace functional, and
\(\mathcal L^+\) is the Drazin inverse of the Liouvillian. With the
vectorization convention
\[
|A\rho B\rangle\!\rangle
=
(B^T\otimes A)|\rho\rangle\!\rangle ,
\]
the left and right response superoperators are
\begin{equation}
\begin{aligned}
\mathcal L_L
&=
-i
\left(
\mathbb I\otimes\partial_\theta H_{\rm eff}
-
(\partial_\theta H_{\rm eff})^T\otimes\mathbb I
\right)
+
\sum_k
\partial_\theta\bar L_k\otimes L_k ,
\\
\mathcal L_R
&=
i
\left(
\mathbb I\otimes\partial_\theta H_{\rm eff}^\dagger
-
(\partial_\theta H_{\rm eff}^\dagger)^T\otimes\mathbb I
\right)
+
\sum_k
\bar L_k\otimes\partial_\theta L_k ,
\end{aligned}
\label{eq:SM_LLR_no_fb}
\end{equation}
where
\(H_{\rm eff}=H-\frac{i}{2}\sum_k L_k^\dagger L_k\).
The two response superoperators describe parameter derivatives acting
on the left and right sides of the trajectory amplitudes.

\subsection{Feedback Liouvillian operator}

We now include instantaneous feedback after each detected jump. A jump
of type \(k\) first produces the unnormalized state
\(L_k\rho L_k^\dagger\), which is then transformed by a completely
positive trace-preserving feedback map \(\mathcal F_k\). The
feedback-controlled master equation is
\begin{eqnarray}
\dot\rho
&=&
\mathcal L^{({\rm fb})}[\rho]\\
&=&
-i[H,\rho]
+
\sum_k
\left(
\mathcal F_k[L_k\rho L_k^\dagger]
-
\frac12\{L_k^\dagger L_k,\rho\}
\right).\nonumber
\label{eq:SM_Lfb}
\end{eqnarray}
The trace-preserving property implies
\(\langle\!\langle\mathbb I|
\overline{\mathcal F}_k
=
\langle\!\langle\mathbb I|\), where
\(\overline{\mathcal F}_k\) is the matrix representation of
\(\mathcal F_k\) in Liouville space. If the feedback maps are unital, one
also has
\(\overline{\mathcal F}_k|\mathbb I\rangle\!\rangle
=
|\mathbb I\rangle\!\rangle\).

With the vectorization convention stated above, the feedback Liouvillian
matrix is
\begin{equation}
\begin{aligned}
\mathbb L^{({\rm fb})}
&=
-i
\left(
\mathbb I\otimes H
-
H^T\otimes\mathbb I
\right)
+
\sum_k
\overline{\mathcal F}_k
(\bar L_k\otimes L_k)
\\
&\quad
-
\frac12
\sum_k
\left[
\mathbb I\otimes L_k^\dagger L_k
+
(L_k^\dagger L_k)^T\otimes\mathbb I
\right].
\end{aligned}
\label{eq:SM_Lmat_fb}
\end{equation}
The feedback steady state
\(|\rho_{\rm ss}^{({\rm fb})}\rangle\!\rangle\) is the normalized
zero-eigenvalue right eigenvector of \(\mathbb L^{({\rm fb})}\), and
\(\mathbb L^{({\rm fb})+}\) denotes the corresponding Drazin inverse on
the subspace orthogonal to the stationary state.

\subsection{Feedback-modified trajectory Fisher information}

Repeating the trajectory Fisher-information derivation for the
feedback-controlled record gives
\begin{equation}
\begin{aligned}
\frac{\mathcal F_{\rm fb}(\theta)}{4\tau}
&=
\sum_k
\left\langle
(\partial_\theta L_k)^\dagger
(\partial_\theta L_k)
\right\rangle_{\rm ss}^{({\rm fb})}
\\
&\quad
-
\langle\!\langle\mathbb I|
\mathcal L_L^{({\rm fb})}
\mathbb L^{({\rm fb})+}
\mathcal L_R^{({\rm fb})}
|\rho_{\rm ss}^{({\rm fb})}\rangle\!\rangle
\\
&\quad
-
\langle\!\langle\mathbb I|
\mathcal L_R^{({\rm fb})}
\mathbb L^{({\rm fb})+}
\mathcal L_L^{({\rm fb})}
|\rho_{\rm ss}^{({\rm fb})}\rangle\!\rangle .
\end{aligned}
\label{eq:SM_FI_fb}
\end{equation}
The feedback-modified response superoperators are
\begin{subequations}
\begin{align}
\mathcal L_L^{({\rm fb})}
&=
-i
\left(
\mathbb I\otimes\partial_\theta H_{\rm eff}
-
(\partial_\theta H_{\rm eff})^T\otimes\mathbb I
\right)
+
\sum_k
\overline{\mathcal F}_k
(\partial_\theta\bar L_k\otimes L_k)
\nonumber\\
&\quad
+
\sum_k
(\partial_\theta\overline{\mathcal F}_k)
(\bar L_k\otimes L_k),
\\[4pt]
\mathcal L_R^{({\rm fb})}
&=
i
\left(
\mathbb I\otimes\partial_\theta H_{\rm eff}^\dagger
-
(\partial_\theta H_{\rm eff}^\dagger)^T\otimes\mathbb I
\right)
+
\sum_k
\overline{\mathcal F}_k
(\bar L_k\otimes\partial_\theta L_k)
\nonumber\\
&\quad
+
\sum_k
(\partial_\theta\overline{\mathcal F}_k)
(\bar L_k\otimes L_k).
\end{align}
\label{eq:SM_LLR_fb}
\end{subequations}
The last terms account for a feedback protocol that itself depends on
the estimated parameter. In the main text we consider a prescribed
feedback strategy, so that
\(\partial_\theta\mathcal F_k=0\). In this case the last terms in
Eq.~\eqref{eq:SM_LLR_fb} vanish.

Equation~\eqref{eq:SM_FI_fb} shows how feedback changes the trajectory
Fisher information. First, the steady state is replaced by
\(\rho_{\rm ss}^{({\rm fb})}\). Second, the feedback maps enter the
propagation of parameter perturbations through
\(\mathcal L_{L,R}^{({\rm fb})}\). Although a trace-preserving feedback
map does not change the instantaneous probability weight of the jump
that triggered it, it changes the conditional post-jump state and
therefore modifies future waiting-time and jump statistics.

When the feedback maps are replaced by the identity channel,
\(\mathcal F_k[\rho]=\rho\), Eq.~\eqref{eq:SM_FI_fb} reduces to the
standard trajectory Fisher information of continuously monitored open
quantum systems. Thus the formula used in the main text is a direct
extension of the feedback-free measurement-record Fisher information to
jump-conditioned feedback trajectories.

The key physical implication is that feedback can enhance or suppress
the parameter sensitivity of the monitoring record by reshaping the
trajectory ensemble. This provides a route to improving trajectory-level
precision without necessarily increasing the reservoir entropy
production, although the feedback apparatus itself carries the
information-thermodynamic cost discussed in
Sec.~\ref{sec:SM_feedback_TUR_cost}.

\section{Derivation of the feedback-modified TUR and feedback entropy cost}
\label{sec:SM_feedback_TUR_cost}

In this section we derive the feedback-modified thermodynamic
uncertainty relation from an enlarged measurement-feedback process and
specify the coarse-grained feedback entropy cost used in the double
quantum dot calculation. The reduced feedback master equation describes
only the controlled system. The thermodynamic cost of operating the
detector, memory, and controller must therefore be assigned within an
enlarged process.

We denote a state of the enlarged dynamics by $x=(s,m,c)$, where
$s$, $m$, and $c$ label the system, memory, and controller degrees
of freedom. An elementary transition is denoted by
$\alpha:x\rightarrow y$, with rate $W_\alpha$, and its reverse
transition by $\alpha^*:y\rightarrow x$, with rate $W_{\alpha^*}$.
The stationary probability of state $x$ is $\pi_x$. For each
forward-backward pair we define
$q_\alpha = \pi_x W_\alpha$,
$q_{\alpha^*} = \pi_y W_{\alpha^*}$,
$j_\alpha = q_\alpha - q_{\alpha^*}$, and
$a_\alpha = q_\alpha + q_{\alpha^*}$. The thermodynamic force on this
edge is
\begin{equation}
A_\alpha = \ln\frac{q_\alpha}{q_{\alpha^*}} .
\end{equation}
The entropy production rate of the enlarged process is then
\begin{equation}
\dot\Sigma_{\rm eff}
= \sum_{\alpha>0} j_\alpha A_\alpha ,
\label{eq:SM_eff_entropy_rate}
\end{equation}
where each pair $(\alpha,\alpha^*)$ is counted once. This effective
entropy production naturally separates into the reservoir contribution
and the feedback-apparatus contribution,
\begin{equation}
\dot\Sigma_{\rm eff}
= \dot\Sigma_{\rm th} + \dot\Sigma_{\rm fb}^{\rm info}.
\end{equation}
The first term is generated by energy and particle exchange with the
thermal reservoirs. The second term is generated by measurement, memory
writing, conditional control, and memory reset.

To obtain the TUR, we introduce an entropic perturbation of the enlarged
Markov process. The perturbed rates are chosen as
\begin{equation}
W_\alpha^\theta
= W_\alpha \exp\!\left[ \theta\,\frac{j_\alpha}{a_\alpha} \right],
\qquad
W_{\alpha^*}^\theta
= W_{\alpha^*} \exp\!\left[ -\theta\,\frac{j_\alpha}{a_\alpha} \right].
\label{eq:SM_entropic_perturbation}
\end{equation}
To first order in $\theta$, this perturbation rescales the irreversible
edge currents while preserving the stationary distribution,
$\partial_\theta j_\alpha^\theta|_{\theta=0} = j_\alpha$.
Hence, for a time-integrated current
$J_\tau[\Gamma] = \sum_\alpha d_\alpha N_\alpha[\Gamma]$, with
$d_{\alpha^*} = -d_\alpha$, one has in the long-time limit
\begin{equation}
\partial_\theta \langle J_\tau\rangle_\theta \big|_{\theta=0}
= \langle J_\tau\rangle + o(\tau).
\label{eq:SM_current_response_entropic}
\end{equation}
The current used in the reduced feedback model is obtained by assigning
nonzero increments only to the corresponding system-reservoir exchange
transitions. Pure detector, memory, and controller transitions carry no
counting increment for this current, but they contribute to the entropy
production of the enlarged process.

\begin{figure}[htb]
\begin{center}
\centering\includegraphics[width=12.5cm]{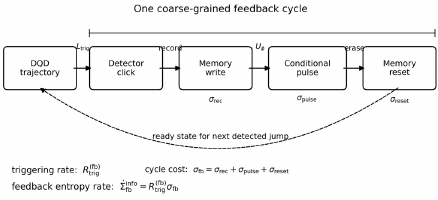}
\caption{Coarse-grained feedback cycle used to assign the information entropy
production of the feedback apparatus. A detected triggering jump
\(L_{\rm trig}\) initiates a feedback cycle consisting of detection,
record writing, conditional control, and memory reset. The entropy cost
of one cycle is denoted by \(\sigma_{\rm fb} =\sigma_{\rm rec} +\sigma_{\rm pulse} + \sigma_{\rm reset}\). Since triggering events occur with stationary rate
\(R_{\rm trig}^{({\rm fb})}\), the corresponding feedback entropy
production rate is \(\dot\Sigma_{\rm fb}^{\rm info} = R_{\rm trig}^{({\rm fb})}\sigma_{\rm fb}\). For the DQD protocol considered in the main text, \(L_{\rm trig}=L_{L,+}\).}
\label{fig:figureSM2}
\end{center}
\end{figure}

The Fisher information of the perturbed trajectory ensemble satisfies
\begin{equation}
\mathcal{I}_\theta
= \tau \sum_{\alpha>0} \frac{j_\alpha^2}{a_\alpha} + o(\tau).
\end{equation}
Using the inequality
$\frac{j_\alpha^2}{a_\alpha}
\le \frac12 j_\alpha \ln\frac{q_\alpha}{q_{\alpha^*}}$,
we obtain
\begin{equation}
\mathcal{I}_\theta
\le \frac{\tau}{2}\,\dot\Sigma_{\rm eff} + o(\tau).
\label{eq:SM_FI_entropy_bound}
\end{equation}
The Cram\'er--Rao bound applied to the trajectory observable $J_\tau$
gives
\begin{equation}
\operatorname{Var}(J_\tau)
\ge \frac{\bigl[ \partial_\theta
           \langle J_\tau\rangle_\theta \big|_{\theta=0} \bigr]^2}
          {\mathcal{I}_\theta}.
\end{equation}
Combining this result with Eqs.~\eqref{eq:SM_current_response_entropic}
and \eqref{eq:SM_FI_entropy_bound}, and taking the long-time limit,
yields
\begin{equation}
\frac{\langle\!\langle J^2\rangle\!\rangle_{\rm fb}}
     {\langle \dot J\rangle_{\rm fb}^2}
\ge \frac{2}{\dot\Sigma_{\rm eff}} .
\label{eq:SM_feedback_TUR}
\end{equation}
This is the feedback-modified TUR used in the main text. The derivation
shows that the relevant entropy production is that of the enlarged
measurement-feedback process, not only the entropy production of the
thermal reservoirs appearing in the reduced system dynamics.

We now specify the coarse-grained feedback cost used in the DQD
calculation. In a microscopic controller model,
$\dot\Sigma_{\rm fb}^{\rm info}$ would be obtained by summing
$j_\alpha A_\alpha$ over all detector, memory, control, and reset
transitions. In the reduced DQD model these internal controller
transitions are not resolved. We therefore assign a total entropy cost
$\sigma_{\rm fb}$ to one complete feedback cycle. Each cycle is
initiated by a feedback-triggering jump, followed by detection, record
writing, conditional control, and memory reset. If
$R_{\rm trig}^{(\mathrm{fb})}$ denotes the total stationary occurrence
rate of the triggering jump channels, then
\begin{equation}
\dot\Sigma_{\rm fb}^{\rm info}
= R_{\rm trig}^{(\mathrm{fb})} \sigma_{\rm fb}.
\label{eq:SM_feedback_entropy_cg}
\end{equation}
For a protocol triggered by a single jump channel,
$R_{\rm trig}^{(\mathrm{fb})}$ reduces to the stationary jump rate of
that channel. In the charge-monitored DQD considered in the main text,
the only triggering channel is the left-injection jump $L_{L,+}$, so
$R_{\rm trig}^{(\mathrm{fb})} = R_{L,+}^{(\mathrm{fb})}$, with
$R_{L,+}^{(\mathrm{fb})}
= \operatorname{tr}[ L_{L,+} \rho_{\rm ss}^{(\mathrm{fb})}
   L_{L,+}^\dagger ]$.

For the minimal controller model used in the numerical calculations, we
write
\begin{equation}
\sigma_{\rm fb}
= \sigma_{\rm mem} + \sigma_{\rm pulse}.
\label{eq:SM_sigma_fb_split}
\end{equation}
The memory part represents the recording and resetting of a binary
feedback record. We take the minimal Landauer-scale cost
$\sigma_{\rm mem} = \ln 2$. The pulse part accounts for the work input
needed to implement the conditional unitary. Immediately after a
left-injection jump the DQD is in the state $\ket{L}$. The feedback
pulse rotates it as
$U_\phi\ket{L}
= \cos\phi\,\ket{L} - i\sin\phi\,\ket{R}$. For
\begin{equation}
H_{\rm DQD}
= \epsilon_L\ket{L}\!\bra{L}
  + \epsilon_R\ket{R}\!\bra{R}
  + t_c \bigl( \ket{L}\!\bra{R} + \ket{R}\!\bra{L} \bigr),
\end{equation}
the energy change induced by this pulse is
\begin{equation}
\Delta E_{\rm pulse}
= \operatorname{tr}\!\bigl[ H_{\rm DQD}\,
   U_\phi\ket{L}\!\bra{L}U_\phi^\dagger \bigr]
  - \bra{L}H_{\rm DQD}\ket{L}
= (\epsilon_R - \epsilon_L) \sin^2\phi .
\label{eq:SM_pulse_energy}
\end{equation}
The tunneling term does not contribute because the coherence generated
by this rotation is purely imaginary. We assign the non-negative entropy
cost
\begin{equation}
\sigma_{\rm pulse}
= \beta_c [\Delta E_{\rm pulse}]_+,
\qquad
[x]_+ = \max(x,0),
\end{equation}
where $\beta_c$ is the inverse temperature of the controller
environment. Possible work extraction for $\Delta E_{\rm pulse} < 0$ is
not credited as a negative cost. Thus the feedback-cycle cost used in
the main text is
\begin{equation}
\sigma_{\rm fb}
= \ln 2 + \beta_c \bigl[ (\epsilon_R - \epsilon_L) \sin^2\phi \bigr]_+ .
\label{eq:SM_sigma_fb_final}
\end{equation}

This $\sigma_{\rm fb}$ should be understood as a minimal coarse-grained
entropy cost per successful feedback cycle. It does not capture the
complete dissipation of a realistic charge detector, amplifier, pulse
generator, and classical controller. A more microscopic apparatus model
would generally contain additional irreversible transitions and would
therefore increase $\dot\Sigma_{\rm fb}^{\rm info}$. Such additional
losses strengthen the feedback-modified TUR. The purpose of the present
coarse-grained construction is to identify the
information-thermodynamic contribution that is absent from the reduced
DQD description and is needed to bound the precision of
feedback-controlled trajectories.

\section{Charge-monitored double quantum dot used in the main text}
\label{sec:SM_DQD_model}

This section collects the explicit generator, counting convention, and
feedback-cost assignment used for the charge-monitored double quantum
dot in the main text. It is intended as a compact reference, not as a
repetition of the physical discussion given there.

\subsection{System Hamiltonian and jump operators}

We consider a spinless double quantum dot in the strong Coulomb blockade
regime. The Hilbert space is spanned by the empty state $\ket{0}$ and
the localized single-electron states $\ket{L}$ and $\ket{R}$. The
system Hamiltonian is
\begin{equation}
H_{\rm DQD}
= \epsilon_L \ket{L}\!\bra{L}
  + \epsilon_R \ket{R}\!\bra{R}
  + t_c \bigl( \ket{L}\!\bra{R} + \ket{R}\!\bra{L} \bigr),
\label{eq:SM_DQD_H}
\end{equation}
where the empty state is taken as the zero of energy, $\epsilon_L$ and
$\epsilon_R$ are the local dot energies, and $t_c$ is the coherent
interdot tunneling amplitude.

The DQD is coupled to two electronic reservoirs $v=L,R$, with inverse
temperatures $\beta_v$, chemical potentials $\mu_v$, and tunneling
rates $\Gamma_v$. The reservoir Fermi function is $f_v(\epsilon) = \frac{1}{e^{\beta_v(\epsilon-\mu_v)}+1}$.
The elementary tunneling jump operators are
\begin{equation}
L_{L,+}
= \sqrt{\Gamma_L f_L(\epsilon_L)}\,\ket{L}\!\bra{0},
\qquad
L_{L,-}
= \sqrt{\Gamma_L[1-f_L(\epsilon_L)]}\,\ket{0}\!\bra{L},
\label{eq:SM_DQD_left_jumps}
\end{equation}
and
\begin{equation}
L_{R,+}
= \sqrt{\Gamma_R f_R(\epsilon_R)}\,\ket{R}\!\bra{0},
\qquad
L_{R,-}
= \sqrt{\Gamma_R[1-f_R(\epsilon_R)]}\,\ket{0}\!\bra{R}.
\label{eq:SM_DQD_right_jumps}
\end{equation}
$L_{v,+}$ describes an electron entering the DQD from reservoir $v$,
and $L_{v,-}$ an electron leaving the DQD into reservoir $v$. The
corresponding environmental entropy increments are
\begin{equation}
\Delta s_{v,+} = -\beta_v(\epsilon_v-\mu_v),
\qquad
\Delta s_{v,-} = +\beta_v(\epsilon_v-\mu_v),
\label{eq:SM_DQD_entropy_increments}
\end{equation}
with $\epsilon_v=\epsilon_L$ for $v=L$ and $\epsilon_v=\epsilon_R$
for $v=R$. These increments implement local detailed balance for the
elementary tunneling events.

\subsection{Jump-conditioned feedback protocol and feedback master equation}

The monitored channel used for feedback is the left-injection jump
$L_{L,+}$. When this jump is detected, the post-jump state is
$\ket{L}$ and the controller applies a fast unitary pulse in the
single-electron subspace,
\begin{equation}
\mathcal{F}_{L,+}[\rho] = U_\phi \rho U_\phi^\dagger,
\qquad
\mathcal{F}_k[\rho] = \rho \quad (k \neq L,+).
\label{eq:SM_DQD_feedback_map}
\end{equation}
The feedback unitary is
\begin{equation}
U_\phi
= \ket{0}\!\bra{0}
  + \exp\!\bigl[ -i\phi( \ket{L}\!\bra{R}
                       + \ket{R}\!\bra{L} ) \bigr],
\label{eq:SM_DQD_feedback_unitary}
\end{equation}
which rotates the single-electron states as
\begin{equation}
U_\phi\ket{L}
= \cos\phi\,\ket{L} - i\sin\phi\,\ket{R},
\qquad
U_\phi\ket{R}
= -i\sin\phi\,\ket{L} + \cos\phi\,\ket{R}.
\label{eq:SM_DQD_feedback_rotation}
\end{equation}
For $\phi\simeq\pi/2$ the injected electron is rotated close to the
right-dot state, increasing the probability of a subsequent output event
into the right reservoir. The map $\mathcal{F}_{L,+}$ is unitary and
therefore completely positive, trace-preserving, and unital.

The feedback-controlled reduced dynamics is
\begin{eqnarray}
\dot\rho
&=& \mathcal{L}^{(\mathrm{fb})}[\rho]\\
&=& -i[H_{\rm DQD},\rho]
  + \sum_{k} \Bigl(
    \mathcal{F}_k[L_k\rho L_k^\dagger]
    - \frac12\{L_k^\dagger L_k,\rho\}
    \Bigr),\nonumber
\label{eq:SM_DQD_QME}
\end{eqnarray}
where $k\in\{L,+,; L,-; R,+,; R,-\}$. The stationary state satisfies
$\mathcal{L}^{(\mathrm{fb})}[\rho_{\rm ss}^{(\mathrm{fb})}]=0$ with
$\operatorname{tr}\rho_{\rm ss}^{(\mathrm{fb})}=1$, and the stationary
rate of jump channel $k$ is
\begin{equation}
R_k^{(\mathrm{fb})}
= \operatorname{tr}\!\bigl[ L_k \rho_{\rm ss}^{(\mathrm{fb})}
   L_k^\dagger \bigr].
\label{eq:SM_DQD_jump_rate}
\end{equation}

\subsection{Counting statistics and clock precision}

In Fig.~3 of the main text, right-reservoir output events are used as
the monitored clock signal. The counting increments are
$c_{R,-}=+1$, $c_{R,+}=-1$, $c_{L,+}=c_{L,-}=0$, so that $R,-$
events correspond to electrons delivered into the right reservoir.
The tilted feedback Liouvillian reads
\begin{equation}
\mathcal{L}_{\chi}^{(\mathrm{fb})}[\rho]
= -i[H_{\rm DQD},\rho]
  + \sum_k \Bigl(
    e^{i\chi c_k}
    \mathcal{F}_k[L_k\rho L_k^\dagger]
    - \frac12\{L_k^\dagger L_k,\rho\}
    \Bigr).
\label{eq:SM_DQD_tilted}
\end{equation}
Let $\lambda_{\rm fb}(\chi)$ be the eigenvalue of
$\mathcal{L}_{\chi}^{(\mathrm{fb})}$ with the largest real part. The
steady current and zero-frequency noise are
\begin{equation}
J_{\rm clk}
= \partial_{i\chi}\lambda_{\rm fb}(\chi)\big|_{\chi=0},
\qquad
S_{\rm clk}
= \partial_{i\chi}^{2}\lambda_{\rm fb}(\chi)\big|_{\chi=0}.
\label{eq:SM_DQD_current_noise}
\end{equation}
Equivalently, $J_{\rm clk} = R_{R,-}^{(\mathrm{fb})}
- R_{R,+}^{(\mathrm{fb})}$. The clock precision rate used in the main
text is $\mathcal{P}_{\rm clk} = J_{\rm clk}^2/S_{\rm clk}$.


The reservoir entropy production rate of the reduced DQD dynamics is
\begin{equation}
\dot\Sigma_{\rm th}
= \sum_k R_k^{(\mathrm{fb})} \Delta s_k .
\label{eq:SM_DQD_Sigma_th_jump}
\end{equation}
Explicitly,
\begin{equation}
\begin{aligned}
\dot\Sigma_{\rm th}
&= R_{L,+}^{(\mathrm{fb})} [-\beta_L(\epsilon_L-\mu_L)]
   + R_{L,-}^{(\mathrm{fb})} [\beta_L(\epsilon_L-\mu_L)] \\
&\quad + R_{R,+}^{(\mathrm{fb})} [-\beta_R(\epsilon_R-\mu_R)]
   + R_{R,-}^{(\mathrm{fb})} [\beta_R(\epsilon_R-\mu_R)] .
\end{aligned}
\label{eq:SM_DQD_Sigma_th_explicit}
\end{equation}
This is equivalent to
$\dot\Sigma_{\rm th}
= -\sum_{v=L,R} \beta_v ( \dot E_v - \mu_v \dot N_v )$,
where $\dot E_v$ and $\dot N_v$ are energy and particle currents from
reservoir $v$ into the system.


The reduced master equation does not include the detector, memory,
amplifier, pulse generator, or reset mechanism. Their thermodynamic
contribution is represented by a coarse-grained entropy cost per
feedback cycle. Since one feedback cycle is initiated whenever the
left-injection jump $L_{L,+}$ is detected,
\begin{equation}
\dot\Sigma_{\rm fb}^{\rm info}
= R_{L,+}^{(\mathrm{fb})} \sigma_{\rm fb}.
\label{eq:SM_DQD_Sigma_fb}
\end{equation}

For the numerical results in Fig.~3 we use the minimal coarse-grained
choice
\begin{equation}
\sigma_{\rm fb}
= \ln 2 + \beta_{\rm ctrl} [ \Delta E_{\rm pulse} ]_+,
\qquad
[x]_+ = \max(x,0).
\label{eq:SM_DQD_sigma_fb_general}
\end{equation}
The term $\ln 2$ is the Landauer-scale cost of recording and resetting
one binary feedback outcome. The second term assigns a non-negative
entropy cost to the conditional pulse; a pulse that lowers the system
energy is not credited as negative dissipation. The pulse energy is
evaluated from the change of the DQD internal energy immediately after
the feedback-triggering jump. Before the pulse the post-jump state is
$\ket{L}$; after the pulse it becomes $U_\phi\ket{L}$. Using
Eq.~\eqref{eq:SM_DQD_feedback_rotation},
\begin{equation}
\Delta E_{\rm pulse}
= \operatorname{tr}\!\bigl[ H_{\rm DQD}\,
   U_\phi\ket{L}\!\bra{L}U_\phi^\dagger \bigr]
  - \bra{L}H_{\rm DQD}\ket{L}
= (\epsilon_R - \epsilon_L)\sin^2\phi.
\label{eq:SM_DQD_pulse_energy}
\end{equation}
The coherent tunneling term does not contribute because the feedback
pulse creates a purely imaginary coherence between $\ket{L}$ and
$\ket{R}$, while the tunneling matrix element is real. Thus
\begin{equation}
\sigma_{\rm fb}
= \ln 2 + \beta_{\rm ctrl}
  \bigl[ (\epsilon_R - \epsilon_L)\sin^2\phi \bigr]_+ .
\label{eq:SM_DQD_sigma_fb_final}
\end{equation}

This expression is a minimal coarse-grained cost, not a microscopic
model of a specific experimental controller. A realistic feedback
apparatus would generally contain additional irreversibility, which
would increase $\dot\Sigma_{\rm fb}^{\rm info}$ and therefore
strengthen the feedback-modified TUR. When $\phi=0$, the feedback
operation reduces to the identity. If the detector and controller
remain active, the binary record still has to be processed and reset,
giving $\sigma_{\rm fb}=\ln2$. If the feedback loop is completely
switched off, then $\dot\Sigma_{\rm fb}^{\rm info}=0$.

The effective entropy production rate entering the feedback-modified
TUR is
\begin{equation}
\dot\Sigma_{\rm eff}
= \dot\Sigma_{\rm th} + \dot\Sigma_{\rm fb}^{\rm info}.
\label{eq:SM_DQD_Sigma_eff}
\end{equation}


The conventional reservoir-only ratio and the feedback-modified ratio
are
\begin{equation}
\mathcal{R}_{\rm th}
= \frac{\dot\Sigma_{\rm th}}{2\mathcal{P}_{\rm clk}}
= \frac{S_{\rm clk}\dot\Sigma_{\rm th}}{2J_{\rm clk}^2},
\qquad
\mathcal{R}_{\rm eff}
= \frac{\dot\Sigma_{\rm eff}}{2\mathcal{P}_{\rm clk}}
= \frac{S_{\rm clk}\dot\Sigma_{\rm eff}}{2J_{\rm clk}^2}.
\label{eq:SM_DQD_TUR_ratios}
\end{equation}
The reservoir-only TUR would require $\mathcal{R}_{\rm th}\ge1$; under
feedback this condition can fail because $\dot\Sigma_{\rm th}$ does not
include the cost of the feedback loop. The feedback-modified TUR
requires $\mathcal{R}_{\rm eff}\ge1$. The comparison between these two
ratios in Fig.~3 demonstrates that the improved clock precision is
bounded only after the information-thermodynamic cost of the controller
is included.

For diagnostic purposes one may define the minimum coarse-grained cost
per triggering event required to restore the bound,
\begin{equation}
\sigma_{\rm fb}^{\min}
= \max\!\left[ 0,\,
   \frac{1}{R_{L,+}^{(\mathrm{fb})}}
   \bigl( 2\mathcal{P}_{\rm clk}
        - \dot\Sigma_{\rm th} \bigr)
   \right].
\label{eq:SM_DQD_sigma_min}
\end{equation}
This quantity is not used as the physical definition of the feedback
cost in Fig.~3; it is only an a posteriori diagnostic. The physical
cost used in the plotted effective ratio is
Eq.~\eqref{eq:SM_DQD_sigma_fb_final}.


The density matrix is vectorized using the column-stacking convention,
${\rm vec}(A\rho B) = (B^T\otimes A)\,{\rm vec}(\rho)$.
For unitary feedback,
${\rm vec}(U_\phi\rho U_\phi^\dagger)
= (U_\phi^*\otimes U_\phi)\,{\rm vec}(\rho)$.
The tilted Liouvillian is a $9\times9$ matrix. The steady state is
obtained by solving the linear system corresponding to
$\mathcal{L}^{(\mathrm{fb})}[\rho_{\rm ss}^{(\mathrm{fb})}]=0$ together
with $\operatorname{tr}\rho_{\rm ss}^{(\mathrm{fb})}=1$. The scaled
cumulant generating function is the eigenvalue of
$\mathcal{L}_\chi^{(\mathrm{fb})}$ with the largest real part.
Derivatives are evaluated by symmetric finite differences,
\begin{equation}
J_{\rm clk}
\simeq {\rm Re}\!\left[
\frac{\lambda_{\rm fb}(h)-\lambda_{\rm fb}(-h)}{2ih}
\right],
\qquad
S_{\rm clk}
\simeq {\rm Re}\!\left[
-\frac{\lambda_{\rm fb}(h)+\lambda_{\rm fb}(-h)-2\lambda_{\rm fb}(0)}{h^2}
\right],
\label{eq:SM_DQD_finite_differences}
\end{equation}
with a small real counting step $h$. The current obtained from this
expression is checked against the direct rate expression
$J_{\rm clk} = R_{R,-}^{(\mathrm{fb})} - R_{R,+}^{(\mathrm{fb})}$.

\section{Bosonic-bath validation in a driven three-level system}
\label{sec:SM_three_level_engine}

\begin{figure}[htb]
\begin{center}
\centering\includegraphics[width=10.5cm]{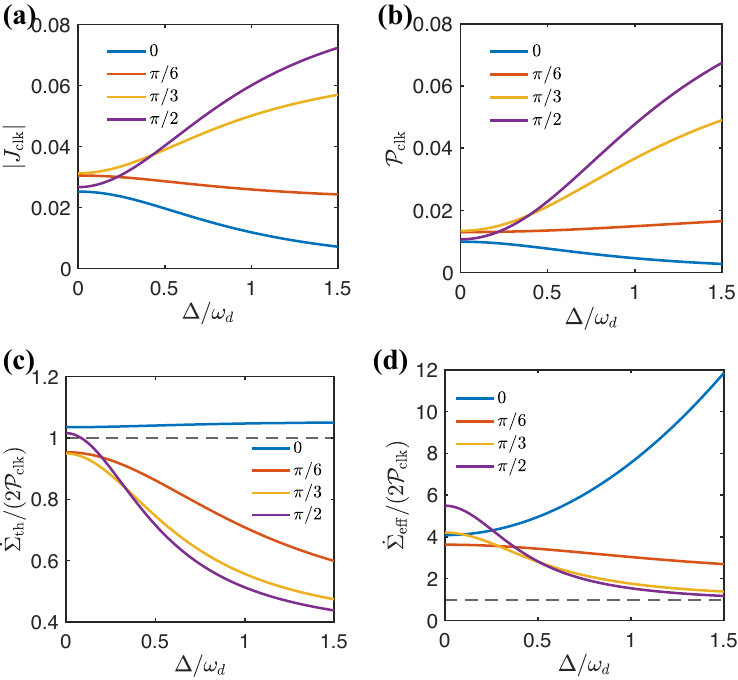}
\caption{
Additional validation of the feedback-modified thermodynamic uncertainty
relation in a driven three-level system.
The system consists of three states $\ket{0}$, $\ket{1}$, and $\ket{2}$.
The hot bath couples $\ket{0}\leftrightarrow\ket{2}$, the cold bath
couples $\ket{1}\leftrightarrow\ket{2}$, and a classical drive couples
$\ket{0}\leftrightarrow\ket{1}$.
Feedback is triggered by the cold-emission jump
$L_{c,-}:\ket{2}\to\ket{1}$, after which the conditional unitary
$U_\phi=\exp[-i\phi(\ket{0}\!\bra{1}+\ket{1}\!\bra{0})]$
rotates the $\{\ket{0},\ket{1}\}$ subspace.
Different curves correspond to $\phi=0,\pi/6,\pi/3,\pi/2$, with
$\phi=0$ giving the no-feedback reference.
The horizontal axis is the normalized drive detuning
$\Delta/\omega_d$, where $\omega_d=\omega_h-\omega_c$.
(a) Net output tick current
$J_{\rm clk}=\dot N_{h,+}-\dot N_{h,-}$, defined from hot-bath
absorption and emission events.
(b) Clock-like precision rate
$\mathcal{P}_{\rm clk}=J_{\rm clk}^2/S_{\rm clk}$, where
$S_{\rm clk}$ is the zero-frequency noise of the same current.
(c) Reservoir-only TUR ratio
$\dot\Sigma_{\rm th}/(2\mathcal{P}_{\rm clk})$.
The dashed horizontal line marks the conventional TUR threshold;
curves with $\phi>0$ can fall below unity, showing that the reservoir
entropy production alone is insufficient to bound the enhanced
precision.
(d) Feedback-modified TUR ratio
$\dot\Sigma_{\rm eff}/(2\mathcal{P}_{\rm clk})$, where
$\dot\Sigma_{\rm eff}
= \dot\Sigma_{\rm th} + \dot\Sigma_{\rm fb}^{\rm info}$.
The feedback entropy production is
$\dot\Sigma_{\rm fb}^{\rm info}
= R_{c,-}^{({\rm fb})}\sigma_{\rm fb}$, with
$R_{c,-}^{({\rm fb})}$ the stationary rate of the
feedback-triggering jump and
$\sigma_{\rm fb}
= \ln 2 + \beta_{\rm ctrl}E_{\rm pulse}\sin^2\phi$.
Including this information-thermodynamic cost restores the TUR, keeping
the effective ratio above unity.
Parameters: $\omega_h=3.0$, $\omega_c=1.8$, $\omega_d=1.2$,
$\varepsilon_2=\omega_h$, $\Omega=1.13$, $\gamma_h=0.255$,
$\gamma_c=0.692$, $\beta_h=0.326$, $\beta_c=0.996$,
$\beta_{\rm ctrl}=\beta_c$, $E_{\rm pulse}=0.404$.
}
\label{fig:figureSM3}
\end{center}
\end{figure}

We further validate the feedback-modified TUR in a driven three-level
system. This example is physically distinct from the
charge-monitored DQD considered in the main text: the monitored jumps
are induced by bosonic heat baths rather than by electronic particle
reservoirs.

The system has three states $\ket{0}$, $\ket{1}$, and $\ket{2}$, with
bare energies $E_0<E_1<E_2$. The hot bath couples the transition
$\ket{0}\leftrightarrow\ket{2}$, the cold bath couples
$\ket{1}\leftrightarrow\ket{2}$, and a classical near-resonant drive
couples $\ket{0}\leftrightarrow\ket{1}$. In the rotating frame and
under the rotating-wave approximation, the effective Hamiltonian is
\begin{equation}
H
= \Delta\ket{1}\!\bra{1}
  + \varepsilon_2\ket{2}\!\bra{2}
  + \frac{\Omega}{2}
    \bigl( \ket{0}\!\bra{1} + \ket{1}\!\bra{0} \bigr),
\label{eq:SM_H_3LS}
\end{equation}
where $\Delta=E_1-E_0-\omega$ is the detuning, $\Omega$ is the Rabi
frequency of the drive, and the energy reference is chosen such that
$E_0=0$.

The heat-bath jump operators are
\begin{equation}
L_{h,+}
= \sqrt{\gamma_h n_h}\,\ket{2}\!\bra{0},
\qquad
L_{h,-}
= \sqrt{\gamma_h(n_h+1)}\,\ket{0}\!\bra{2},
\label{eq:SM_hot_jumps_3LS}
\end{equation}
for the hot bath and
\begin{equation}
L_{c,+}
= \sqrt{\gamma_c n_c}\,\ket{2}\!\bra{1},
\qquad
L_{c,-}
= \sqrt{\gamma_c(n_c+1)}\,\ket{1}\!\bra{2},
\label{eq:SM_cold_jumps_3LS}
\end{equation}
for the cold bath. Here
$n_h = [e^{\beta_h\omega_h}-1]^{-1}$,
$n_c = [e^{\beta_c\omega_c}-1]^{-1}$,
with $\omega_h=E_2-E_0$ and $\omega_c=E_2-E_1$. The feedback-free
dynamics is generated by
\begin{equation}
\dot\rho
= -i[H,\rho]
  + \sum_{a=h,c}\sum_{\nu=\pm}
    \Bigl( L_{a,\nu}\rho L_{a,\nu}^\dagger
         - \frac12\{L_{a,\nu}^\dagger L_{a,\nu},\rho\} \Bigr).
\end{equation}

We monitor the heat-bath jumps and apply feedback after the
cold-emission jump $L_{c,-}$. This jump brings the system from $\ket{2}$
to $\ket{1}$, i.e., to the upper state of the driven transition. The
feedback pulse is a unitary rotation in the
$\{\ket{0},\ket{1}\}$ subspace,
\begin{equation}
\mathcal{F}_{c,-}[\rho]
= U_\phi \rho U_\phi^\dagger,
\qquad
\mathcal{F}_k[\rho] = \rho \quad (k \neq c,-),
\label{eq:SM_feedback_3LS}
\end{equation}
where
\begin{equation}
U_\phi
= \exp\!\bigl[ -i\phi( \ket{0}\!\bra{1}
                       + \ket{1}\!\bra{0} ) \bigr].
\label{eq:SM_Uphi_3LS}
\end{equation}
For $\phi=0$ the feedback is absent, while for $\phi=\pi/2$ the pulse
maps $\ket{1}$ to $-i\ket{0}$, conditionally accelerating the
completion of the engine cycle after a cold-bath emission event.

The feedback-controlled Liouvillian is obtained by replacing the
$L_{c,-}\rho L_{c,-}^\dagger$ jump branch with
$\mathcal{F}_{c,-}[L_{c,-}\rho L_{c,-}^\dagger]$. The tilted feedback
generator is constructed as in the main text by multiplying the
corresponding monitored jump branches by $e^{i\chi d_k}$. For the
hot-bath heat current, one chooses $d_{h,+}=+\omega_h$,
$d_{h,-}=-\omega_h$, and $d_k=0$ otherwise.

The thermal entropy production rate is
\begin{equation}
\dot\Sigma_{\rm th}
= -\beta_h \dot Q_h - \beta_c \dot Q_c ,
\end{equation}
where $\dot Q_a$ is the heat current flowing into reservoir $a$. The
feedback contribution is assigned using the same coarse-grained
controller model as in the main text,
\begin{equation}
\dot\Sigma_{\rm fb}^{\rm info}
= R_{c,-}^{(\mathrm{fb})} \sigma_{\rm fb},
\label{eq:SM_fb_cost_3LS}
\end{equation}
where
$R_{c,-}^{(\mathrm{fb})}
= \operatorname{tr}[ L_{c,-} \rho_{\rm ss}^{(\mathrm{fb})}
   L_{c,-}^\dagger ]$
is the stationary rate of the feedback-triggering cold-emission jumps.
The effective entropy production entering the TUR is
$\dot\Sigma_{\rm eff}
= \dot\Sigma_{\rm th} + \dot\Sigma_{\rm fb}^{\rm info}$.

We use this model as an additional numerical check of the
feedback-modified TUR. Feedback reduces the reservoir-only TUR ratio
$\dot\Sigma_{\rm th}/(2\mathcal{P}_J)$, which may fall below unity. The
effective ratio $\dot\Sigma_{\rm eff}/(2\mathcal{P}_J)$, however,
remains above unity once the feedback information cost is included.
This confirms that the feedback-modified TUR applies to
feedback-controlled bosonic jump processes in a driven system, and
is not restricted to the electronic DQD transport setting discussed in
the main text.

Figure~\ref{fig:figureSM3} provides an independent numerical
validation of the feedback-modified TUR in a driven three-level
system. Unlike the charge-monitored DQD of the main text, where the
monitored output events are fermionic tunneling jumps, here the ticks are
bosonic heat-bath jumps. The agreement between the two examples
demonstrates that restoring the TUR through the effective entropy
production is not specific to electronic transport, but a generic
consequence of embedding the measurement-feedback apparatus in the
thermodynamic description.

The horizontal axis is the normalized detuning \(\Delta/\omega_d\) of the
driven \(\{\ket{0},\ket{1}\}\) subspace, which acts as an effective
feedback-controlled qubit. Varying \(\Delta\) shifts the working point of
this qubit while keeping the hot- and cold-bath transition frequencies
fixed, so that the observed changes reflect primarily the interplay
between the coherent drive, the conditional feedback pulse, and the
thermal jump statistics. The four curves correspond to feedback angles
\(\phi=0,\pi/6,\pi/3,\pi/2\). The case \(\phi=0\) gives the identity
dynamics, while finite \(\phi\) rotates the state in the
\(\{\ket{0},\ket{1}\}\) subspace after a detected cold-emission jump
\(L_{c,-}\).

Figure~\ref{fig:figureSM3}(a) shows the magnitude of the output tick current
\(J_{\rm clk} = \dot N_{h,+} - \dot N_{h,-}\), which counts net hot-bath
absorption events. These events mark the repeated operation cycles of the
system and serve as clock-like ticks, in analogy to the output electron
jumps used in the DQD clock of the main text. Without feedback, the
current decreases as the detuning grows, indicating that moving the
driven qubit away from its optimal working point suppresses the net
output. Finite feedback reverses this trend: for \(\phi=\pi/3\) and
\(\phi=\pi/2\) the current is strongly enhanced over a broad detuning
range. Physically, the cold-emission jump prepares the system in
\(\ket{1}\), and the feedback pulse rotates this state toward
\(\ket{0}\), thereby accelerating the hot-bath absorption branch of the
thermodynamic cycle.
Figure~\ref{fig:figureSM3}(b) shows the corresponding clock-like precision rate
\(\mathcal{P}_{\rm clk} = J_{\rm clk}^2/S_{\rm clk}\). The no-feedback
curve decreases with detuning, reflecting both a weaker signal and a loss
of regularity. Finite feedback not only sustains but enhances
\(\mathcal{P}_{\rm clk}\) at large detuning, showing that the conditional
pulse improves both the average tick rate and the timing regularity of
the output record.

Figure~\ref{fig:figureSM3}(c) tests whether this enhanced precision can still be bounded by
the reservoir entropy production alone. The quantity plotted is
\(\dot\Sigma_{\rm th}/(2\mathcal{P}_{\rm clk})\), where
\(\dot\Sigma_{\rm th}\) contains only the entropy production associated
with heat exchange in the hot and cold baths. The dashed horizontal line
marks the conventional TUR threshold. The no-feedback curve remains
above unity, consistent with the standard bound. The feedback curves,
however, fall below unity for sufficiently strong \(\phi\), confirming
that the reservoir entropy production is no longer the complete cost of
the feedback-stabilized clock precision. The reason is clear:
\(\dot\Sigma_{\rm th}\) accounts for the thermal reservoirs but ignores
the thermodynamic cost of detecting the trigger jump, writing the
measurement record, generating the conditional pulse, and resetting the
controller memory.  
Figure~\ref{fig:figureSM3}(d) restores the bound by including the feedback information
entropy production \(\dot\Sigma_{\rm fb}^{\rm info}\). The effective
entropy production \(\dot\Sigma_{\rm eff} = \dot\Sigma_{\rm th}
+ \dot\Sigma_{\rm fb}^{\rm info}\) is evaluated using the same
coarse-grained model as in the main text:
\(\dot\Sigma_{\rm fb}^{\rm info} = R_{c,-}^{(\mathrm{fb})}\sigma_{\rm fb}\),
with a cycle cost \(\sigma_{\rm fb} = \ln 2
+ \beta_{\mathrm{ctrl}}E_{\rm pulse}\sin^2\phi\). The first term is the
Landauer-scale cost of recording and resetting a binary feedback outcome;
the second term accounts for the non-negative thermodynamic effort of
producing a conditional pulse of strength \(\phi\). The resulting
effective TUR ratio \(\dot\Sigma_{\rm eff}/(2\mathcal{P}_{\rm clk})\)
remains above unity for all feedback angles and detunings shown. This
confirms that the apparent violation in panel (c) is resolved once the
information-processing cost of the feedback loop is properly accounted
for, strengthening the universal character of the feedback-modified TUR
derived in the main text.

For \(\phi=0\), the feedback operation is the identity. In panel (d) we
nevertheless include the monitoring and memory-reset cost when the
feedback loop is kept active. If the controller is completely switched
off, this cost vanishes and the \(\phi=0\) effective ratio coincides with
the reservoir-only ratio.

\end{document}